\documentclass[12pt]{article}
\usepackage[margin=2cm]{geometry}
\usepackage{comment}
\usepackage{amsmath,amssymb,extarrows,graphicx,subfigure,setspace}
\usepackage{cite}
\usepackage{slashed}
\usepackage{color}
\makeatother

\usepackage{epsfig,multicol,bbm}

\newcommand\fverb{\setbox\pippobox=\hbox\bgroup\verb}
\newcommand\fverbit{\egroup\item[\fbox{\unhbox\pippobox}]}
\newcommand{\bea}{\begin{eqnarray}}
\newcommand{\eea}{\end{eqnarray}}
\newcommand{\ba}{\begin{array}}
\newcommand{\ea}{\end{array}}
\newcommand{\ee}{\end{equation}}

\newbox\pippobox
\begin{document}
\title{\bf Title should be added}

\begin{titlepage}
	\vspace{10mm}
	\begin{flushright}
		
	\end{flushright}
	
	\vspace*{20mm}
	\begin{center}
		
		{\Large {\bf  More on Complexity in Finite Cut Off Geometry}\\
		}
		
		\vspace*{15mm}
		\vspace*{1mm}
	{S. Sedigheh Hashemi$^{a}$, Ghadir Jafari $^{a}$, Ali Naseh$^{a}$ and Hamed Zolfi$^{a,b}$ 
	}
		
		\vspace*{1cm}
		
		{\it 
			${}^a$ School of Particles and Accelerators,
			Institute for Research in Fundamental Sciences (IPM)\\
			P.O. Box 19395-5531, Tehran, Iran
			\\  
			${}^b$ Department of Physics, Sharif University of Technology,\\
			P.O. Box 11365-9161, Tehran, Iran
		}
		
		\vspace*{0.5cm}
		{E-mails: {\tt ghjafari,hashemiphys,naseh@ipm.ir,\ hamed.Zolfi@physics.sharif.edu}}
		
		\vspace*{1cm}
	\end{center}
	
\begin{abstract}
It has been recently proposed that late time behavior of holographic complexity in a uncharged black brane solution of Einstein-Hilbert theory with boundary cut off is consistent with Lloyd's bound if we have a cut off behind the horizon. Interestingly, the value of this new cut off is fixed by the boundary cut off.  In this paper, we extend this analysis to the charged black holes. Concretely, we find the value of this new cut off for  charged small black hole solutions of Einstein-Hilbert-Maxwell theory, in which the proposed bound on the complexification is saturated. We also explore this new cut off in Gauss-Bonnet-Maxwell theory.
	
\end{abstract}
	
\end{titlepage}

\section{Introduction}

In recent years, amazing and interesting connections have been discovered between theoretical quantum information theory and  quantum gravity. These connections have been provided through AdS/CFT duality by which certain quantities in dual quantum field theory can be related to certain quantities in the bulk spacetime. The leading example of such relation is the Ryu-Takayanagi proposal which provides a geometrical realization of entanglement entropy in a dual CFT \cite{Ryu:2006bv}.
The next example is the Susskind proposal \cite{Susskind:2014moa} by which quantum computational complexity of a boundary state is dualized to a special portion of spacetime. Interestingly enough, this proposal can be used to understand the rich geometric structures that exist behind the horizon and also to construct the dual operators that describe the interior of a black hole. In this way one can, hopefully, resolve the black hole information paradox which has been a focus of attention over the past years \cite{Mathur:2009hf,Almheiri:2012rt,Almheiri:2013hfa,Marolf:2013dba,Papadodimas:2012aq,Papadodimas:2013jku,deBoer:2018ibj}.
This essential property of complexity comes from this fact that even after the boundary theory reached thermal equilibrium it continues to increase. 
\\

The complexity of a quantum target state (systems of qubits) is essentially defined as the minimum number of gates one needs to act on a certain reference state to produce approximately the desired target state \cite{Arora:2009,Moore: 2011}. In this notion, the gates are unitary operators which can be taken from some universal set. For instance, one may be asked how hard it is to prepare the ground state of a local Hamiltonian, which generally is an entangled state, from a reference state which is spatially factorisable. This suggests that there might exists a deep relation between complexity of preparing a quantum state and the entanglement between degrees of freedom in that quantum state.
\\

To describe the quantum complexity of states in boundary QFT, two holographic proposal have been developed, $\text{complexity= volume (CV)}$ conjecture \cite{Susskind:2014moa,Alishahiha:2015rta} and $\text{complexity=action (CA)}$ conjecture \cite{Brown:2015bva,Brown:2015lvg}. In the CA conjecture, which is the  main focus of this manuscript, the quantum complexity of a state, $\mathcal{C}_{A}$, is given by
\bea
\mathcal{C}_{A} = \frac{I_{\text{WDW}}}{\pi \hbar},
\eea
where $I_{\text{WDW}}$ is the on-shell gravitational action evaluated on a certain subregion of spacetime known as the Wheeler-DeWitt (WDW) patch and we will set $\hbar = 1$ from now on. It is worth noting that the intersection of WDW patch with the future interior determines completely the late time behavior of complexity growth rate \cite{Alishahiha:2018lfv} which is in agreement with Lloyd's bound \cite{Lloyd}. The Lloyd's bound says that the upper bound on rate of complexity growth ( for uncharged systems) is twice the energy of  system and this energy in the holographic set up is actually the mass of black hole. According to the holographic calculations at late times
\bea
\frac{d \mathcal{C}_A}{dt} = 2 M_{\text{BH}},
\eea
which is independent of the boundary curvature and the spacetime dimension\footnote{For non-relativistic models see \cite{Swingle:2017zcd,Alishahiha:2018tep}.}. Several aspects of this holographic proposal have been explored. In
particular, its time dependence \cite{Brown:2015bva,Brown:2015lvg,Lehner:2016vdi,Carmi:2017jqz,Yang:2016awy}, the structure of divergences \cite{Carmi:2016wjl,Reynolds:2016rvl}   and
its reaction to shockwaves \cite{Stanford:2014jda,Chapman:2018dem,Chapman:2018lsv}. However, these research programs are in the beginning steps because we still do not have a precise definition of quantum complexity in QFTs.
Some initial steps towards developing such a precise definition have been taken in the recent years \cite{Jefferson:2017sdb,Chapman:2017rqy,Khan:2018rzm,Hackl:2018ptj,Alves:2018qfv,Camargo:2018eof,Caputa:2017urj,Sinamuli:2019utz,Bhattacharyya:2018bbv,Yang:2018tpo,Ali:2018fcz}.\\

Apart from those interesting developments, a new research program has been started to answer this important question: What is a general structure of an effective QFT for which the UV behavior is not described by a CFT and can we holographically probe this field theory?. From the QFT side this question has been answered  by Smirnov and Zamolodchikov \cite{Zamolodchikov:2004ce}. They discovered a general class of exactly solvable irrelevant deformations of two dimensional CFTs \cite{Smirnov:2017,Cavaglia:2016oda}. QFTs typically are connected to UV fixed points (CFTs) by relevant or marginal operators and turning on irrelevant operator spoils the existence of a UV fixed point. It is worth mentioning that irrelevant couplings also typically eradicate locality at some high cut off scale which in itself challenges the UV completeness of these theories. The holographic dual of Smirnov and Zamolodchikov uncharged QFTs are proposed in \cite{McGough:2016lol}. The proposed dual is three dimensional Einstein-Hilbert gravity at finite radial cut off. Soon after that this holographic dual has been explored further (in the same dimensions or higher dimensions, in absence or presence of charges) \cite{Dubovsky:2017cnj,Shyam:2017znq,Kraus:2018xrn,Cardy:2018sdv,Aharony:2018vux,Dubovsky:2018bmo,Taylor:2018xcy,Hartman:2018tkw}. 
\\


Recently in \cite{Akhavan:2018wla}, the holographic complexity in a finite cut off uncharged black brane solutions of Einstein-Hilbert theory is calculated. It is found that the late time behavior of complexity growth rate is consistent with the LIyd's bound if in addition to the finite boundary cut off, there exists another cut off behind the horizon. The consequences of this new cut off were explored more in \cite{Alishahiha:2018swh,Alishahiha:2019cib} and by that the problem of holographic complexity for Jackiw-Teitelboim (JT) gravity is resolved. JT model \cite{Jackiw:1984je,Teitelboim:1983ux} emerges as the holographic description of Sachdev-Ye-Kitaev (SYK) model \cite{Sachdev:92,Kitaev:15} in a particular low energy limit \cite{Maldacena:2016upp}. The SYK model is a strongly coupled quantum many-body system
that is nearly conformal, exactly solvable and chaotic. This unique combination of properties put the SYK model at cutting edge in both high energy and condensed matter physics.   
\\

It is worth mentioning that initial calculations of holographic complexity using CA proposal in JT gravity produced an unexpected result: the growth rate of complexity vanishes at late
time. It is the sign that if JT gravity is a correct holographic dual of SYK model something is missed. Two independent resolutions for this problem is proposed. The first one is based on adding a new boundary term (involving the Maxwell field) to the gravitational action and changing the variational principle \cite{Brown:2018bms,Goto:2018iay} and second one, as we said above, is based on considering a new cut off surface behind the horizon without changing the variational principle \cite{Alishahiha:2018swh,Alishahiha:2019cib}.  
\\

The aim of this paper is to further study the cut off surface behind the horizon in the proposal \cite{Akhavan:2018wla}. To be more precise, we find the relation between this new cut off and boundary finite cut off in charged black hole solutions of both Einstein-Hilbert-Maxwell theory and Gauss-Bonnet-Maxwell theory. We should emphasize again that the only concrete result, which is available in literature for quantum complexity in interacting QFTs comes from holographic calculations. A generalization of Lloyd's bound for charged black holes (U(1) charged systems)
has been proposed in \cite{Brown:2015lvg}. According to this proposal, the natural bound for states at a finite chemical potential is given by
\bea\label{bound}
\frac{d \mathcal{C}_{A}}{dt} \leq \frac{2}{\pi} \bigg[(M-\mu Q)-(M-\mu Q)_{\text{gs}}\bigg],
\eea
where the subscript $gs$ indicates the state of lowest $(M -\mu Q)$ for a given chemical potential $\mu$. The situation for intermediate size ($r_{+} \sim L $) and large charged black holes ($r_{+} \gg
L$) is complicated and leads to an apparent violation of the complexification bound (\ref{bound}) \cite{Carmi:2017jqz}. Fortunately, when we apply this formula to charged black holes with spherical horizon, that are much smaller than the
AdS radius ($r_{+} \ll L$), the rate of change of complexity at late times saturates the bound  
\bea\label{cdotq}
\frac{d \mathcal{C}_{A}}{dt} = \frac{2}{\pi} (M-\mu Q),
\eea
where for a given $\mu$, the smallest value of $(M-\mu Q)$ is zero. This observation 
is the spotlight in our study in the next sections. By comparing the rate of change of complexity at late times in a finite cut off geometry and comparing the result with the right hand side of (\ref{cdotq}) , but with $M$  substituted with the quasi local energy at finite cut off 
we can find the relation between behind the horizon cut off and the boundary cut off. 
\\

It is also worth mentioning that another bound for the growth rate of complexity for $U(1)$ charged system is proposed in \cite{Cai:2016xho}
\bea
\frac{d \mathcal{C}_{A}}{dt} \leq \frac{2}{\pi} \bigg[(M - \mu Q)_{+}-(M-\mu Q)_{-}\bigg],
\eea
where the subscripts $\pm$ indicates
the outer and inner horizons and the chemical potentials are defined as $\mu_{\pm} = Q/r_{\pm}^{d-2}$. One should bear in mind that the chemical potential $\mu_{-}$ has no corresponding quantity at the boundary. Moreover, the only dependency to the mass cancels in subtraction of two terms. This cancellation implies that for all black hole sizes, the growth rate of complexity in a geometry without boundary cut off is the same as one in geometry with that cut off. The consequence of assuming this bound on behind the horizon cut off is studied in \cite{Alishahiha:2019cib}. Although it might be the case that the complexity of a state in a CFT and its irrelevant deformation finally would be the same but as we mentioned previously, calculation of this quantity in QFTs especially for interacting ones is not provided so far. So, in the following we are studying the effect of bound (\ref{cdotq}) on the location of behind the horizon cut off.
\\

The remainder of this paper is organized as follows. In section.\ref{section2} we calculate the holographic complexity for charged black holes in Einstein-Hilbert-Maxwell theory in presence of boundary cut off and behind the horizon cut off. In  small black hole limit, we compare this complexity with $2(E-\mu Q)$ at finite cut off which it gives the relation between those two cut offs. In section.3 we present the relations between those two cut offs in presence of Gauss-Bonnet term. The ingredients to find this relation will be provided in two subsections \ref{QLEGB} and \ref{CGB}. In subsection \ref{QLEGB} we find  quasi local energy $E$ for a finite cut off RN black hole of Gauss-Bonnet-Maxwell  theory. In subsection \ref{CGB} we find holographic complexity for this RN black hole in presence of boundary cut off and behind the horizon cut off. The last section contains a summary and discussion of the main results.  
\section{Einstein-Hilbert-Maxwell Theory at Finite Cut Off}\label{section2}
\begin{figure}
	\begin{center}
		\includegraphics[width=0.4\linewidth]{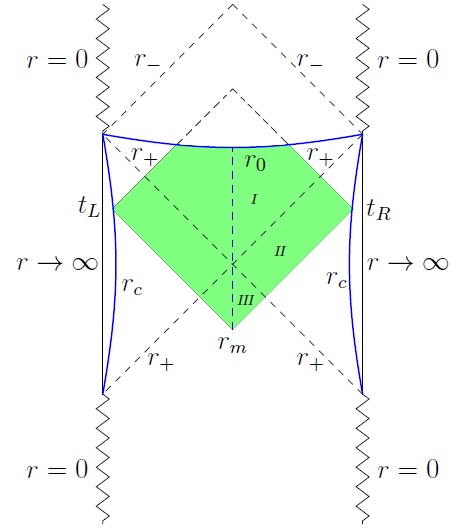}
		\caption{Penrose diagram of the eternal charged AdS black hole. The theory is defined at a radial finite cut off $r_c$ and the actual WDW patch is shown in green color.}\label{Fig1}
	\end{center}
\end{figure}

The holographic complexity of $(d+1)$ dimensional charged AdS black holes in Einstein-Hilbert-Maxwell theory have been studied in \cite{Brown:2015lvg}, however, in  this section we study the same problem but in presence of boundary cut off $r=r_c$ and also $r=r_0$ cut off behind the outer horizon, which they are shown in figure.\ref{Fig1}. We should emphasize that in \cite{Alishahiha:2019cib}  similar calculations are done but for (near extremal) black branes and also $r=r_0$ cut off behind the inner horizon and near to the curvature singularity.\\

The metric of charged AdS black hole solution of Einstein-Hilbert-Maxwell theory takes the general form
\begin{equation}\label{metric}
	{\rm}ds^2=-f(r) {\rm}dt^2+\frac{{\rm d}r^2}{f(r)}+r^2 {\rm d}\Sigma^2_{\kappa, d-1},
\end{equation}
where the blackening factor $f(r)$ is given by
\begin{equation}
	f(r)= \kappa+\frac{r^2}{L^2}-\frac{\omega^{d-2}}{r^{d-2}}+\frac{q^2}{r^{2(d-2)}}.
\end{equation}
Here $L$ is the AdS curvature scale and  $\kappa$ describes the curvature of the $(d-1)$ dimensional line element ${\rm d}\Sigma^2_{\kappa, d-1}$. Black holes with  $k=\{+1, 0, -1\}$ respectively have spherical, planar, and hyperbolic horizon geometries.  The thermodynamic quantities describing the black hole (\ref{metric}) are
\begin{equation}
	M= \frac{(d-1)\Omega_{k,d-1}}{16\pi G}\hspace{1mm} \omega^{d-2},\hspace{.5cm}S = \frac{\Omega_{k,d-1}}{4 G}r_{+}^{d-1},\hspace{.5cm} T= \frac{1}{4\pi} \frac{\partial f}{\partial r}\bigg{|}_{r=r_+}.
\end{equation}
The charge and the Maxwell potential also respectively are \cite{Chamblin:1999tk}
\begin{align}\label{chargemu}
	Q= \oint * F= \frac{q\hspace{1mm} \Omega_{k,d-1}\sqrt{(d-1)(d-2)}}{g \sqrt{8\pi G}},
	\cr \nonumber\\
A_{t}(r) = \frac{g}{\sqrt{8\pi G}} \sqrt{\frac{d-1}{d-2}}\hspace{1mm}\left(\frac{q}{r_+^{d-2}}-\frac{q}{r^{d-2}}\right).
\end{align}
The causal structure of the charged black holes (\ref{metric}) is illustrated by Penrose diagram in figure.\ref{Fig1}. On the two asymptotic boundaries, the constant time slices are denoted by $t_{L}$ and $t_{R}$ and the actual WDW patch is shown in green color. According to the boost symmetry the evaluation of the complexity will depend on $t=t_{L}+t_{R}$ and not on each of the boundary times separately. So in the following we focus on symmetric times $t_{L} = t_{R}=t/2$, without loss of generality. In order to consider the null sheets bounding the WDW patch the tortoise coordinate is defined as
\begin{equation}
	r^{*}(r) = -\int ^{\infty}_{r}  \frac{{\rm d}r}{f(r)}, 
\end{equation}
which by that Eddington-Finkelstein coordinates, $u$ and $v$, where describe out- and in-going null rays, become
\bea\label{EFcoordinates}
v= t+r^{*}(r),\hspace{.5cm}u=t-r^{*}(r).
\eea
It is also useful to fix the notation for null vectors respectively associated with constant $v$ and $u$ surfaces
\bea
k_{1} = \alpha \left(\partial_{t}+\frac{1}{f(r)}\partial_{r}\right),\hspace{1cm}k_{2} =\alpha\left(\partial_{t}-\frac{1}{f(r)}\partial_{r}\right),
\eea
here $\alpha$ is constant parameter appearing due to ambiguity of the normalization of null vectors.
\\

The Einstein-Hilbert-Maxwell gravitational action can be written as \cite{Parattu:2015gga,Parattu:2016trq,Lehner:2016vdi}
\begin{equation}\label{THEEACTION}
	\begin{split}
		I& =  \frac{1}{16 \pi G} \int_\mathcal{M} d^{d+1} x \sqrt{-g} \left(R + \frac{d(d-1)}{L^2}-\frac{1}{4 g^2}F^2\right) \\
		&+\frac{1}{8\pi G} \int _{\Sigma_{t}^{d}} K_{t}\hspace{1mm} d\Sigma_{t}\pm\frac{1}{8\pi G} \int _{\Sigma_{s}^{d}} K_{s}\hspace{1mm} d\Sigma_{s}\pm\frac{1}{8\pi G} \int _{\Sigma_{n}^{d}} K_{n}\hspace{1mm} dSd\lambda
		\\
		&
		\pm\frac{1}{8\pi G} \int_{J^{d-1}}  a\hspace{1mm} dS,
	\end{split}
\end{equation}
where the first line contains standard Einstein-Hilbert-Maxwell action including the Ricci scalar $R$, the cosmological constant $\Lambda = -d (d-1)/2 L^2$ plus the electromagnetic field strength tensor $F_{\mu \nu}$, with a coupling constant $g$.
The second line contains the Gibbons-Hawking-York (GHY) terms which are needed to have a well-defined variational principle respectively on timelike, spacelike and null boundaries. The $K_{i}$'s denote the extrinsic curvatures and $\lambda$ is the null coordinate defined on the null segments. The user manual for the sign of different terms in (\ref{THEEACTION}) can be found in \cite{Lehner:2016vdi}. At last, the function $a$ is defined at the intersection of two null boundaries and it is given by
\bea\label{a}
a= \log{\frac{\mid \hspace{-1mm}k_1.k_2\hspace{-1mm}\mid}{2}}.
\eea
In what follows, we will consider the action calculation on the WDW patch in the static black hole background (\ref{metric}) with the causal structure shown in figure.\ref{Fig1}, which has three contributions: bulk integration, GHY contribution for the cutt off surface $r_0$ and joint term at $r_m$,
\begin{equation}
	I_{\text{tot}} = I_{\text{bulk}}+I_{\text{GHY}}+I_{\text{joint}}. 
\end{equation}
The reason for ommiting another joint contributions is that they do not contribute to the growth rate of complexity. Moreover, as we use the affine parametrization to parametrize  the null directions, the boundary terms on null segments have zero contribution.\\

To calculate the contribution from the bulk action, we divide the WDW patch into three regions: I, the region between $r_0$ and the outer horizon $r_+$; II, the region outside the outer horizon $r_+$; and finally III, the region behind the outer horizon (see figure.\ref{Fig1}). First we write the integrand in the bulk action as \cite{Cano:2018aqi}
\begin{equation}
	I_{\text{bulk}} = \int _{\text{WDW}} {\rm d}^{d+1}x \sqrt{-g}\hspace{1mm}\mathcal{L}=\int _{\text{WDW}} {\rm d}t\,{\rm d}r \hspace{1mm}I(r),
\end{equation}
in which $I(r) ={\rm d} \mathcal{I}(r)/{\rm d} r$, and 
\begin{equation}\label{I}
	\mathcal{I}(r) = \frac{\Omega_{k,d-1}}{16\pi G }\left[-\frac{2(d-1) q^2}{r^{d-2}}+(d-1) \omega^{d-2} -r^{d-1} f'{}(r)\right].
\end{equation} 
The bulk contributions are given by
\begin{equation}
	\begin{split}
		I^{\text{I}}_{\text{bulk}} = & \, 2 \int^{r_{+}}_{r_{0}}dr\hspace{1mm}I(r)\left(\frac{t}{2}-r^{*}(r)\right),
		\\
		I^{\text{II}}_{\text{bulk}}=&\,4  \int^{r_{\text{max}}}_{r_{+}}dr\hspace{1mm} I(r)\left(-r^{*}(r)\right),
		\\
		I^{\text{III}}_{\text{bulk}} =&\,  2 \int^{r_{+}}_{r_{m}}dr\hspace{1mm} I(r)\left(-\frac{t}{2}-r^{*}(r)\right),
	\end{split}
\end{equation}
where an extra factor of two was added in order  to account for the two sides of the Penrose diagram in figure.\ref{Fig1}.
Adding the above contributions and take a time derivative we arrive to
\begin{equation}
	\frac{{\rm d} I_{\text{bulk}}}{{\rm d}t}= \int ^{ r_{m}}_{r_{0}} I(r)\hspace{1mm}{\rm d}r =  \mathcal{    I        }  (r)\bigg|^{r_m}_{ r_0}.
\end{equation}
Now, substituting (\ref{I}) in the above equation yields
\begin{equation}\label{bulkk}
	\frac{{\rm d}I_{\text{bulk}}}{{\rm d}t}=\frac{\Omega_{k,d-1}}{8\pi G }\left[q^2 \left(\frac{1}{r_{0}^{d-2}}-\frac{1}{r_{m}^{d-2}}\right)+\frac{1}{L^2}\left(r_{0}^d-r_{m}^d\right)\right].
\end{equation}
The GHY action at the cut off surface  $r=r_{0}$ is    
\begin{equation}
	I_{\text{GHY}}= -\frac{2}{8 \pi G}\int_{r=r_0} \hspace{-2mm}{\rm d}t\hspace{.5mm}{\rm d}^{d-1}x \hspace{.5mm} \sqrt{-h}\hspace{.5mm}K_{s},
\end{equation}
where the extrinsic curvature $K$ is given by
\begin{equation}
	K = \frac{n_r}{2}\left(\partial_{r} f(r)+\frac{2(d-1)}{r}f(r)\right),
\end{equation}
and $n_{r}$ is the normal vector to this cut off surface. Consequently, we get 
\begin{equation}
	I _{\text{GHY}}= -\frac{\Omega_{k,d-1}}{8\pi G }r^{d-1} \left(\partial_{r} f(r)+\frac{2 (d-1) f(r)}{r}\right)\left(\frac{t}{2}-r^*(r)\right) \bigg| _{r=r_{0}}.
\end{equation}
Taking the time derivative of $I _{\text{GHY}}$ leads to 
\begin{equation}\label{ghy}
	\frac{{\rm d} I _{\text{GHY}}}{{\rm}dt} \bigg| _{r=r_{0}}\hspace{-2mm}=-\frac{\Omega_{k,d-1}}{16\pi G }\left[\frac{2 q^2}{r_{0}^{d-2}}+\frac{2 (d-1) \kappa}{r_{0}^{2-d}}+\frac{2d	}{L^2}r_{0}^d-d\, \omega^{d-2}\right].
\end{equation}
The time dependent joint term contribution can be calculated using (\ref{a}) and it is given by 
\begin{equation}\label{jointEH}
	I_{\text{joint}}=-\frac{\Omega_{k,d-1}}{8 \pi G}\hspace{1mm}r_m^{d-1}\log\left(\frac{|f(r_m)|}{\alpha^2}\right),
\end{equation}
here $\alpha$ is an arbitrary constant used for the normalization
of the null vectors. Using the identity $
\frac{{\rm d} r_{m}}{{\rm d}t} = - \frac{f(r_{m})}{2}$,
the time derivative of the joint action (\ref{jointEH}) becomes
\begin{equation}\label{jnt}
	\frac{{\rm d}I_{\text{joint}}}{{\rm d}t}=\frac{\Omega_{k,d-1}}{16 \pi G}\left[\frac{2 r_{m}^d}{L^2}-\frac{2 q^2(d-2)}{r_{m}^{d-2}}+(d-2)\omega^{d-2}+\log\frac{|f(r_{m})|}{\alpha^2}f(r_{m})(d-1)r_{m}^{d-2}
	\right].
\end{equation}
To remove the ambiguity associated with the normalization of null vectors, another boundary term should be added to the action
\bea\label{ambiguityy}
I_{\text{amb}} = \frac{1}{8\pi G} \int d\lambda d^{d}x\sqrt{\sigma}\hspace{1mm}\Theta \log\frac{\mid\hspace{-1mm}\tilde{l}\Theta\hspace{-1mm}\mid}{d}, 
\eea
where the induced metric on the joint point is $\sigma$, $\tilde{l}$ is an undetermined length scale and 
\bea\label{Theta}
\Theta = \frac{1}{\sqrt{\sigma}}\frac{\partial \sqrt{\sigma}}{\partial\lambda}.
\eea
Since the late time behavior of growth rate of complexity is important for us and for these times the last term in (\ref{jnt}) vanishes, we do not need to add the ambiguity term (\ref{ambiguityy}) in the following.   
\\

\hspace{-.6cm}Now by summing (\ref{bulkk}), (\ref{ghy}), (\ref{jnt}) and taking the late time limit we find
\begin{equation}\label{cdotEH}
	\dot{\mathcal{C}}=\frac{{\rm d}I_{\text{tot}}}{{\rm d}t}= \frac{(d-1)\Omega_{k,d-1}}{8\pi G }\left[-\frac{r_0^{d}}{L^2}+\omega^{d-2}-\kappa\, r_{0}^{d-2}-\frac{q^2}{r_{+}^{d-2}}\right].
\end{equation}
Note that by substituting $r_0$ with $r_{-}$ in last result, which is equal to $r_c\hspace{-1mm}\rightarrow\hspace{-1mm}\infty$, we recover the well-known result in \cite{Brown:2015lvg,Carmi:2017jqz}. At this point one may ask whether we should add contribution of standard boundary counterterms given by
\bea\label{ctEH}
I_{\text{ct}}= -\frac{1}{16 \pi G}\int _{r=r_0}\hspace{-2mm} {\rm d}^dx \sqrt{-h} \left(\frac{2(d-1)}{L}+\frac{L}{(d-2)}\hspace{1mm}\mathcal{R} +\frac{L^3}{(d-2)^2(d-4)}\hspace{1mm}(\mathcal{R}_{ij}^2-\frac{d}{4(d-1)}\mathcal{R}^2)+...\hspace{-1mm}\right)
\eea
on the $r=r_0$ surface, to the complexity growth rate (\ref{cdotEH}). In the chargless limit $q^2\rightarrow 0$, and also $r_c\rightarrow \infty$ which describes the undeformed theory, the contributions of the above counterterms to  $\dot{\mathcal{C}}$ at late times will diverge or become a constant. Actually these divergent terms do not appear in the same limit in (\ref{cdotEH}) and the constant one apparently violates the LIoyd's bound. Accordingly, in the following we will not consider the counterterm action (\ref{ctEH}).
\\

Now, as mentioned in the introduction, for small charged black holes, $r_-< r_{0}< r_+\ll L$, with spherical horizon one has
\bea\label{cdot}
\dot{\mathcal{C}}= 2 \bigg((E-E_{\text{global}}) -\mu Q\bigg),
\eea
at late times. In presence of  boundary cut off, $E$ and $E_{\text{global}}$ are respectively proportional to the gravitational quasi-local energy of black hole solution and global AdS solution and also $\mu$ is chemical potential at this boundary cut off. In the limit $r_c \rightarrow \infty$, (\ref{cdot}) exactly matches with (\ref{cdotq}). The gravitational quasi-local energy is\footnote{In comparison with Eq.(5.17) of \cite{Hartman:2018tkw}, the third term in Eq.(\ref{EgrEH}) is new. This extra term actually appears just for $d \geq 5$ dimensions. The detail analysis for derivation of (\ref{EgrEH}) is provided in subsection.\ref{QLEGB}.} \cite{Hartman:2018tkw} 
\bea\label{EgrEH}
E_{\text{gr}}= \frac{(d-1)\Omega_{k,d-1}\hspace{1mm}r_{c}^{d-1}}{8\pi G }\left(\frac{1}{L}+\frac{\kappa\, L}{2 r_{c}^2 }-\frac{L^3}{8r_c^4}-\sqrt{\frac{1}{L^2}-\frac{\omega^{d-2}}{r_{c}^d}
+\frac{q^2}{r_{c}^{2d-2}}+\frac{\kappa}{r_{c}^2}}\hspace{2mm}\right),
\eea
and its relation with $E$ is given by
\bea\label{E}
E_{\text{gr}} = \frac{L}{r_c} E.
\eea
Substituting back (\ref{EgrEH}) in (\ref{cdot}) and using (\ref{E}) and (\ref{chargemu}) gives\footnote{Note that $\mu \equiv A_{t}(r_c)$.}
\bea\label{ee}
&& \hspace{-.8cm}2 \left((E-E_{\text{global}})- \mu Q\right)=
\cr \nonumber\\
&& \frac{2(d-1)\Omega_{k,d-1}}{8\pi G }\left[\frac{r_{c}^d}{L}\left(\sqrt{\frac{1}{L^2}+\frac{\kappa}{r_{c}^2}}-\sqrt{\frac{1}{L^2}-\frac{\omega^{d-2}}{r_{c}^d}
+\frac{q^2}{r_{c}^{2d-2}}+\frac{\kappa}{r_{c}^2}}\hspace{2mm}\right)
-q^2\left(\frac{1}{r_{+}^{d-2}}-\frac{1}{r_{c}^{d-2}}\right)\right].\nonumber\\ 
\eea
To proceed further one can simplify  (\ref{ee}) more by noting that for small black holes $q^2 = r_{-}^{d-2}r_{+}^{d-2}$. Using this simplified version of (\ref{ee}) and equating it
with  (\ref{cdotEH}), at leading order in $r_c$, we arrive to 
\bea\label{r0EH}
r_0= r_{-}\left(1 + \frac{L^2}{2(d-2)r_c^2}\hspace{1mm}(1+\frac{r_+^{d-2}}{r_{-}^{d-2}})\right).
\eea
Before closing this section, it is worth emphasizing that the conjectured relation (\ref{cdotq}) is known to fail in the intermediate times \cite{Brown:2015lvg,Carmi:2017jqz}.
The obtained relations in (\ref{r0EH}) are reliable in the late times and violation of generalized Lloyd's bound just modifies these relations in the way that $r_0$ becomes a function of $t$. Finding this time dependancy is not in the scope of this work.
\section{Gauss-Bonnet-Maxwell Theory at Finite Cut Off}
In this section we first find the energy-momentum tensor at finite radial cut off for Gauss-Bonnet-Maxwell theory and by that we calculate the quasi local energy. After that we calculate the complexity growth rate at finite cut off geometry. By having these two ingredients we can find the relation between boundary cut off $r_c$ and behind the outer horizon cut off $r_0$.
\subsection{Quasi local energy at finite cut off}\label{QLEGB}
To find the quasi local energy the  standard method is the Brown-York Hamilton-Jacobi prescription. In this approach one uses derivative of the action with respect to the induced metric on the timelike boundary. One of the prerequisites in this approach is that the gravitational action should have a well-posed variational principle. Another requirement is that we should add local counterterms to the boundary action, so that the energy for a reference spacetime vanishes. Accordingly, the total renormalized action is
\begin{equation}
	I_{\text{ren}}= I_{\text{bulk}}+I_{\text{GHY}}- I_{\text{ct}}.
\end{equation}
The gravitational stress-tensor (which equivalently can be interpreted as expectation value of boundary field theory stress-tensor) becomes
\begin{equation}\label{T}
T^{ij} = \frac{2}{\sqrt{-h}}\frac{\delta I_{\text{ren}}}{\delta h_{ij}}=2\, \pi ^{ij} -2\,  P^{ij},
\end{equation}
with
\begin{equation}\label{piP}
\pi ^{ij} = \frac{1}{\sqrt{-h}}\frac{\delta (I_{\text{bulk}}+I_{\text{GHY}})}{\delta h_{ij}},\hspace{1cm}P^{ij} = \frac{1}{\sqrt{-h}}\frac{\delta I_{\text{ct}}}{\delta h _{ij}},
\end{equation}
and $h_{ab}$ is induced metric on the timelike boundary. The bulk action of Gauss-Bonnet-Maxwell theory is
\begin{equation}\label{bulk}
I_{\text{bulk}}=  \frac{1}{16 \pi G} \int _{\mathcal{M}}{\rm d}^{d+1}x \sqrt{-g}\hspace{1mm}\bigg( R+\frac{d(d-1)}{L^2}-\frac{1}{4 g^2}F_{\mu \nu}^2+\alpha_{\text{GB}} \left(R^2-4 R_{\mu \nu}^2 + R_{\mu \nu \rho \sigma}^2\right)\bigg),
\end{equation}
where $\alpha_{\text{GB}}$ is the Gauss-Bonnet coefficient with dimension $\text{(length)}^2$. Moreover, the proper GHY term is \cite{Myers:1987yn}
\bea\label{GBBT}
I_{\text{GHY}}=\frac{1}{16\pi G}\int _{\partial \mathcal{M}} {\rm d}^dx \sqrt{-h} \hspace{1mm}\bigg(2K +\alpha_{\mathrm{GB}}\bigg[ 8 \hspace{.5mm}\mathcal{G}_{ij} K^{ij} - \tfrac{8}{3} K_{i}{}^{j} K^{i k} K_{k j} + 4 K K_{ij} K^{ij} -  \tfrac{4}{3} K^3\bigg]\bigg),
\eea
and the proper counterterm, just for small value of $\alpha_{\mathrm{GB}}$,  becomes \cite{Liu:2008zf}
\bea\label{ct}
I_{\text{ct}}= -\frac{1}{16 \pi G}\int _{\partial \mathcal{M}} {\rm d}^dx \sqrt{-h} \left(I_{\text{ct}}^{(0)}+I_{\text{ct}}^{(2)} \hspace{1mm}\mathcal{R} +I_{\text{ct}}^{(4)}\hspace{1mm}(\mathcal{R}_{ij}^2-\frac{d}{4(d-1)}\mathcal{R}^2)+...\right),
\eea
with
\bea
&& I_{\text{ct}}^{(0)} = \frac{2(d-1)}{L}\left(1- \frac{\alpha_{\mathrm{GB}} (d-2)(d-3)}{6L^2}\right),
\cr \nonumber\\
&& I_{\text{ct}}^{(2)} =\frac{L}{(d-2)}\left(1+ \frac{3\alpha_{\mathrm{GB}}(d-2)(d-3)}{2 L^2}\right) ,
\cr \nonumber\\
&&I_{\text{ct}}^{(4)} = \frac{L^3}{(d-2)^2(d-4)}\left(1- \frac{15\alpha_{\mathrm{GB}}(d-2)(d-3)}{2L^2}\right).\nonumber
\eea
The $\mathcal{G}_{ij}$, $K_{ij}$ and $ \mathcal{R}, \mathcal{R}_{ij} $ are respectively Einstein tensor, extrinsic curvature tensor and intrinsic curvature tensors of induced metric $h_{ij}$. It is worth noting that in (\ref{ct}) the standard logarithmic counterterm action is absent. The reason is that for static spherical black hole solutions of the model (\ref{bulk}),   
\bea\label{key}
	ds^2=-f(r)dt^2+\frac{1}{f(r)}dr^2+r^2 d\Sigma_{{\kappa, d-1}},
\eea
with
\bea\label{fGB}
	f(r)= \kappa +\frac{r^2}{2 \tilde{\alpha}_{\mathrm{GB}}}\left[1 - \sqrt{1+4 \tilde{\alpha}_{\mathrm{GB}}\left(\frac{\omega^{d-2}}{r^d}-\frac{1}{L^2}-\frac{q^2}{r^{2(d-1)}}\right)}\hspace{2mm}\right],\hspace{.5cm}\tilde{\alpha}_{\mathrm{GB}}= \alpha_{\mathrm{GB}} (d-2)(d-3) 
\eea
the contribution of this counterterm to $P^{ab}$ is zero. From the stress-energy tensor (\ref{T}), the energy surface density $\varepsilon$ is defined by the normal projection of $T^{ij}$ on a co-dimension two surface ($r,t=\text{cte}$),
\begin{equation}
\varepsilon = T^{tt} u_{t} u_{t} =  2 \pi^{tt} u_{t} u_{t}-  2 P^{tt} u_{t} u_{t},
\end{equation}
and the total quasi local energy is given by
\bea\label{Egr}
E_{\text{gr}}= \int {\rm d}^{d-1}x \sqrt{\sigma}\, \varepsilon,
\eea
where $\sigma$ is the metric on that co-dimension two surface. Using (\ref{piP}) and (\ref{bulk})-(\ref{ct}), one can see that
\bea
&&2 \pi^{tt} u_{t} u_{t}=  (d-1)\frac{2\sqrt{f(r)}}{r}\left(-1 - \frac{2\tilde{\alpha}_{\mathrm{GB}}}{r^2}  (\kappa- \frac{1}{3}f(r)\hspace{.1mm}) \right),
\cr \nonumber\\
&& 2 P^{tt} u_{t} u_{t}= -(d-1)\left( \frac{2}{L}+\frac{\kappa L}{r^2}-\frac{L^3}{4r^4}-\frac{\tilde{\alpha}_{\text{GB}}}{L^2}(\hspace{.2mm}\frac{1}{3L}
-\frac{3\kappa L }{2 r^2}-\frac{15 L^3}{8r^4}\hspace{.2mm})\hspace{.2mm} \right),
\eea
and the quasi local energy (\ref{Egr}) at finite cut off $r=r_c$ becomes
\bea
&& E_{\text{gr}} =\frac{(d-1)\Omega_{\kappa, d-1}\,r_c^{d-1}  }{16 \pi G} \bigg[\frac{2}{L}+\frac{\kappa L}{r_c^2}-\frac{L^3}{4r_c^4}- \frac{2}{r_c}\sqrt{f(r_c)}
\cr \nonumber\\
&&\hspace{3.5cm}
-\frac{\tilde{\alpha}_{\text{GB}}}{L^2}( \frac{1}{3L}
-\frac{3\kappa L}{2 r_c^2}-\frac{15 L^3}{8r_c^4})
-4\tilde{\alpha}_{\text{GB}} \frac{\sqrt{f(r_c)}}{r_c^3}(\kappa- \frac{1}{3}f(r_c) )\bigg].
\eea 
To translate this energy to the energy in boundary field theory, one should note that
\bea\label{EgrGB}
E_{\text{gr}} = \int {\rm d}^{d-1}x \sqrt{\sigma}\hspace{1mm}T_{tt}\hspace{1mm} u^{t}u^{t}= \int {\rm d}^{d-1}x\hspace{1mm}( \frac{r^{d-1}}{L^{d-1}}\sqrt{\tilde{\sigma}})\hspace{.5mm}(\frac{r^{2-d}}{L^{2-d}} \tilde{T}_{tt})\hspace{.5mm}(\frac{L^2}{r^2}\tilde{g}^{tt}) = \frac{L}{r_c} E,
\eea
where the field theory energy is 
\bea
E= \int {\rm d}^{d-1}x\sqrt{\tilde{\sigma}}\tilde{T}_{tt}\,  \tilde{g}^{tt}.
\eea
Consequently, we have
\bea\label{EGB}
&& E =\frac{(d-1)\Omega_{\kappa, d-1}\,r_c^{d}}{16 \pi G L} \hspace{1mm}\bigg[\frac{2}{L}+\frac{\kappa L}{r_c^2}-\frac{L^3}{4r_c^4}- \frac{2}{r_c}\sqrt{f(r_c)}
\cr \nonumber\\
&&\hspace{3.5cm}
-\frac{\tilde{\alpha}_{\text{GB}}}{L^2}( \frac{1}{3L}
-\frac{3\kappa L}{2 r_c^2}-\frac{15 L^3}{8r_c^4})
-4\tilde{\alpha}_{\text{GB}} \frac{\sqrt{f(r_c)}}{r_c^3}(\kappa- \frac{1}{3}f(r_c) )\bigg].
\eea 
\subsection{Complexity in Gauss-Bonnet-Maxwell theory at finite cut off}\label{CGB}
In this subsection we find the holographic complexity for charged AdS black hole solution (\ref{key}) at finite cut off . The steps are the same as section.\ref{section2}.  It should be noted that for general values of Gauss-Bonnet coupling, the causal structure of the charged black hole (\ref{key}) differs from figure.\ref{Fig1}. In order to avoid a singularity before the inner horizon a sufficient condition is to demand $0 \leq \alpha_{\mathrm{GB}} < \frac{L^2}{4}$.
\\

First, we consider the contribution from the bulk action. Instead of $\mathcal{I}(r)$ in Einstein-Hilbert theory (\ref{I}), for Gauss-Bonnet-Maxwell theory one has \cite{Cano:2018aqi}
\begin{multline}
\mathcal{I}(r) = \frac{\Omega_{k,d-1}}{16\pi G }\left[-\frac{2(d-1) q^2}{r^{d-2}}+(d-1) \omega^{d-2}
-r^{d-1} f'{}(r)\left(1-2\tilde{\alpha}_{\text{GB}} \left(\frac{f(r)-\kappa}{r^2}\right)\frac{d-1}{d-3}\right)\right],
\end{multline}
which leads to
\begin{multline}\label{bulkGB}
\frac{{\rm d}I_{\text{bulk}}}{{\rm d}t} = \frac{\Omega_{k,d-1}}{16\pi G }\left[2 q^2(d-1)(\frac{1}{r_{0}^{d-2}}-\frac{1}{r_{m}^{d-2}})-r_{m}^{d-1} f\rq{}(r_{m})\left(1-2\tilde{\alpha}_{\text{GB}} \left(\frac{f(r_m)-\kappa}{r_m^2}\right)\frac{d-1}{d-3}\right)\right. \\ \left.+r_{0}^{d-1} f\rq{}(r_{0})\left(1-2\tilde{\alpha}_{\text{GB}} \left(\frac{f(r_0)-\kappa}{r_0^2}\right)\frac{d-1}{d-3}\right)
\right].
\end{multline}
Let us now consider the GHY term (\ref{GBBT}) on spacelike $r=r_0$ surface. After a short computation one can see that
\begin{multline}\label{GHGB}
\frac{{\rm d}I_{\text{GHY}}}{{\rm d}t}\bigg|_{r = r_0}\hspace{-.35cm}=\frac{(d-1)\Omega_{\kappa, d-1}}{16 \pi G}\left[
-2 r_0^{d-2} f(r_0)-\frac{1}{d-1}r_0^{d-1} f'(r_0)\right.\\
\left.
	-4\tilde{\alpha}_{\text{GB}}\left( r_0^{d-4}f(r_0)\big(\kappa-\frac{1}{3}f(r_0)\big)+\frac{1 }{2(d-3)}r_0^{d-3}f'{}(r_0)\big(\kappa- f(r_0)\big)
	\right)\right].
\end{multline}
Next, we consider the contribution from the joint term at $r=r_m$. Another joint contributions are not time dependent. The joint term for intersections of null boundaries is given by
\cite{jafari2019} 
\bea\label{Joint}
&& I_{\text{joint}} = \pm\frac{1}{8 \pi G} \int d^{d-1}x\sqrt{\sigma} \hspace{1mm}\log{\frac{\mid k_1.k_2\mid}{2}},
\cr \nonumber\\
&&\hspace{1.3cm}	\pm\frac{1}{4\pi G} \int d^{d-1}x\sqrt{\sigma} \bigg[\mathcal{R}[\sigma]\log{\frac{\mid k_1.k_2\mid}{2}}+\frac{4}{k_1.k_2}\big(\Theta^{(1)}_{ab}\Theta^{(2)\hspace{.1mm}ab}-\Theta^{(1)\hspace{.1mm}a}_{a}\Theta^{(2)\hspace{.1mm}b}_{b}\big)\bigg]\alpha_{\mathrm{GB}},
\eea
where
\bea
\Theta^{(i)}_{ab} = \frac{1}{2} \partial_{\lambda_{i}}\sigma_{ab},
\eea
and $\mathcal{R}[\sigma]$ is Ricci-scalar curvature of the joint metric. This boundary action for black hole solution (\ref{key}) becomes 
\bea
I_{\text{joint}}=-\frac{\Omega_{k,d-1}}{8 \pi G}\hspace{1mm}\bigg[r_m^{d-1}\log\left(\frac{|f(r_m)|}{\alpha^2}\right)\left(1+2\kappa\frac{\tilde{\alpha}_{\text{GB}}}{r_m^2} \left(\frac{d-1}{d-3}\right)\right)+4\tilde{\alpha}_{\text{GB}} \left(\frac{d-1}{d-3}\right)\frac{f(r_m)}{r_m^2}\bigg],
\eea
and it's time derivative is given by
\bea\label{jointt}
&& \hspace{-3mm}\frac{{\rm d}I_{\text{joint}}}{{\rm d}t}=\frac{\Omega_{k,d-1}}{16 \pi G}\left(r_m^{d-1}f^{\prime}(r_m)+\log\frac{|f(r_{m})|}{\alpha^2}f(r_{m})(d-1)r_{m}^{d-2}
\right)\left(1+2\kappa\frac{\tilde{\alpha}_{\text{GB}}}{r_m^2} \left(\frac{d-1}{d-3}\right)\right)
\cr \nonumber\\
&&\hspace{.9cm}-\kappa\hspace{1mm}\tilde{\alpha}_{\text{GB}}\frac{ \Omega_{k,d-1}}{4 \pi G}\left(\frac{d-1}{d-3}\right) r_m^{d-4}\log\left(\frac{|f(r_m)|}{\alpha^2}\right)f(r_m)  +\tilde{\alpha}_{\mathrm{GB}}\frac{\Omega_{k,d-1}}{4 \pi G}\left(\frac{d-1}{d-3}\right)\left(\frac{f(r_m)}{r_m^2}\right)^{\prime}f(r_m).\nonumber\\
\eea
At late times, $r_m \rightarrow r_+$, the ambiguity terms and also the last term in second line vanish. Therefore, just for the late times we have 
\bea\label{jointGB}
\frac{{\rm d}I_{\text{joint}}}{{\rm d}t} =\frac{\Omega_{k,d-1}}{16 \pi G}\hspace{1mm} r_{m}^{d-1} f\rq{}(r_{m})\left(1+2\kappa\frac{\tilde{\alpha}_{\text{GB}}}{r_m^2} \left(\frac{d-1}{d-3}\right)\right).
\eea
Finally, let us  show that the time derivative of the null boundary terms vanish. For given a null segment  parametrized by $\lambda$ and
with a transverse space metric $\sigma_{ab}$, the boundary contribution will have the form \cite{Chakraborty:2018dvi, jafari2019}
\bea\label{nullGB}
&& I_{\text{null}} =\frac{1}{8 \pi G} \int d\lambda d^{d-1}x \sqrt{\sigma} \hspace{.5mm}K_{n} 
\cr \nonumber \\
&&\hspace{.9cm} -\frac{\tilde{\alpha}_{\text{GB}}}{16\pi G} \int d\lambda d^{d-1}x \sqrt{\sigma}\hspace{1mm}\bigg[
-8\hspace{.5mm} \Xi_{\nu}^{\mu} \hspace{.5mm}\mathcal{L}_{k}\Theta_{\mu}^{\nu}
+ 8 \hspace{.5mm}K_{n}\hspace{.5mm} \Theta_{\nu}^{\mu}\hspace{.5mm}\Xi_{\mu}^{\nu}-8\hspace{.5mm}\Theta_{\rho}^{\mu}\hspace{.5mm}\Theta_{\nu}^{\rho}\hspace{.5mm}\Xi_{\mu}^{\nu}-8K_{n} \Theta \hspace{.5mm}\Xi+8 \hspace{.5mm}\Xi \hspace{.5mm}\mathcal{L}_{k}\Theta \cr \nonumber\\
&&\hspace{3.5cm}+8\hspace{.5mm}\Xi\hspace{.5mm} \Theta_{\nu}^{\mu}\hspace{.5mm}\Theta^{\nu}_{\mu}
-4 K_{n} \mathcal{R}[\sigma]+16(D^{\mu}\Theta-D^{\nu}\Theta_{\nu}^\mu)W_{\mu}+8(\Theta \sigma^{\mu\nu}-\Theta^{\mu\nu})W_{\mu}W_{\nu} \bigg],\nonumber\\
\eea
with $\Theta_{\mu\nu} = \nabla_{\mu}k_{\nu}$, $\Xi_{\mu\nu} = \nabla_{\mu}n_{\nu}$ and 
\bea
&&\Theta_{a b}= P_{ab}^{\mu\nu}\hspace{.5mm} \Theta_{\mu\nu},\hspace{.5cm}
\Xi_{a b}=P_{ab}^{\mu\nu}\hspace{.5mm} \Xi_{\mu\nu},
\hspace{.5cm}W_{\mu} =n^{\nu} \nabla_{\mu}k_{\nu},
\cr \nonumber\\
&&\Theta =\sigma^{ab}\Theta_{a b},\hspace{1cm}\Xi=\sigma^{ab}\Xi_{a b},\hspace{1cm} D^{\mu} = e^{\mu}_{a}D^{a},\hspace{1cm}k^{\mu}n_{\mu}=-1,
\eea
where $P^{\mu\nu}_{ab}$ is projection to the $\sigma_{ab}$ surface. Moreover, $\mathcal{L}_{k}A_{\mu\nu...}$ denotes the Lie derivative of tensor $A_{\mu\nu...}$ tensor along the vector $k^{\mu}$. For affine parameterization of null surface and static black hole (\ref{key}), the boundary term (\ref{nullGB}) can be simplified as
\bea\label{Inull}
I_{\text{null}} =-\frac{(d-1)(d-2)}{\pi G} \hspace{1mm}\alpha_{\mathrm{GB}} \int d\lambda d^{d-1}x \big(1-\alpha f(r)\big) r^{d-4},
\eea
where again $\alpha$ denotes the ambiguity in defining the null generator of a null surface. From (\ref{Inull}), for null boundaries of WDW patch in black hole solution (\ref{key}), at late times one can see that
\bea
\frac{d I_{\text{null}}}{dt} = 0.
\eea
Adding bulk integration (\ref{bulkGB}), boundary contribution (\ref{GHGB}), and joint term (\ref{jointGB}) we will have
\bea\label{cdotGB}
&&\mathcal {\dot{C}}=\frac{{\rm d}}{{\rm d}t}(I_{\text{tot}}+I_{\text{GH}}+I_{\text{joint}}+I_{\text{amb}})
\cr \nonumber\\
&&\hspace{.32cm}= \frac{2 (d-1) \Omega_{k,d-1}}{16 \pi G}\left[\frac{q^2}{r_{0}^{d-2}}-\frac{q^2}{r_{m}^{d-2}}- f(r_0)r_{0}^{d-2}- 2 \tilde{\alpha}_{\text{GB}}\hspace{1mm} f(r_{0}) r_{0}^{d-4}\left(\kappa - \frac{1}{3}f(r_0)\right)\right].
\eea
It is worth noting that in the limit $r_0\rightarrow r_-$, which is equal to $r_c\rightarrow \infty$, the above result at late times exactly matches with the known result \cite{Cano:2018aqi}. Furthermore, because of the same reasons for Einstein-Hilbert theory we do not consider the contribution of counterterm action (\ref{ct}) on $r=r_0$ surface.
\\

Now, same as section.\ref{section2}, in order to obtain the relation between the boundary cut off $r_c$ and finite cut off $r_0$, we set
\bea\label{cdotgbe}
\dot{\mathcal{C}}= 2 \bigg((E-E_{\text{global}})-\mu Q\bigg),
\eea
where $\mathcal {\dot{C}}$ and $E$ are given by (\ref{cdotGB}) and (\ref{EGB}),  respectively, 
and also from \cite{Cano:2018aqi} we have
\bea\label{muGB}
\mu = \frac{g}{\sqrt{8\pi G}} \sqrt{\frac{d-1}{d-2}}\hspace{1mm}\left(\frac{q}{r_+^{d-2}}-\frac{q}{r_c^{d-2}}\right),
\hspace{.75cm}Q = \frac{q\hspace{1mm} \Omega_{k,d-1}\sqrt{(d-1)(d-2)}}{g \sqrt{8\pi G}}.
\eea
Moreover, $E_{\text{global}}$ can be obtained from (\ref{EGB}) by substituting $(\omega^{d-2},q^2=0)$ in (\ref{fGB}). To this aim what we need is the small black hole limit in this theory and its relation with $q^2$. For obtaining this, we note that the blackening factor $f(r)$ in (\ref{key}) satisfies the following constraint
\begin{equation}
h\left(\frac{L^2(f(r)-k)}{r^2}\right) =\frac{\omega^{d-2}L^2}{r^d}-
\frac{q^2 L^2}{r^{2(d-1)}},
\end{equation}
where the polynomial function $h(x)$ is 
\begin{equation}
h(x)= 1-x+\frac{\tilde{\alpha}_{\text{GB}}}{L^2}x^2.
\end{equation}
By defining
\begin{equation}
h_{\pm}=h \left(-\kappa\frac{L^2}{r_{\pm}^2}\right),
\end{equation}
it is easy to see that
\begin{equation}\label{qgb}
q^2= \frac{r_+^d h_+- r_-^d h_-}{r_{+}^{d-2}-r_{-}^{d-2}}\frac{r_{+}^{d-2}r_{-}^{d-2}}{L^2}.
\end{equation}
Now, the small black hole limit  
$r_-, r_+ \ll  L$, implies that
\begin{equation}\label{qtwo}
q^2 = \left(\kappa+\tilde{\alpha}_{\text{GB}}\,\frac{r_{+}^{d-4}-r_{-}^{d-4}}{r_{+}^{d-2}-r_{-}^{d-2}}\right)r_{+}^{d-2}r_{-}^{d-2}.
\end{equation}
Furthermore, the outer horizon is located at the zero of (\ref{fGB}), that is $f(r_{+})=0$, which up to first order in Gauss-Bonnet coupling yields 
\bea\label{omeg}
\omega^{d-2} = \frac{r_{+}^d}{L^2}+\kappa \,r_{+}^{d-2} +\frac{q^2}{r_{+}^{d-2}}+\tilde{\alpha}_{\text{GB}}\,r_{+}^{d-4}.
\eea
 Now we substitute $E$ and $\mu,Q$ respectively from (\ref{EGB}) and (\ref{muGB}) in the right hand side of (\ref{cdotgbe}) and  simplify  the result by using (\ref{qtwo}) and (\ref{omeg}). By equating obtained simplied expression with $\dot{\mathcal{C}}$ given by (\ref{cdotGB}), for near extremal small black holes, $r_{-} \approx r_{+} \ll L$ , and with spherical horizons we find
\bea\label{r0GB}
r_0= r_{-}\left(1 + \frac{L^2}{2(d-2)r_c^2}\hspace{1mm}(1+\frac{r_+^{d-2}}{r_{-}^{d-2}})\right).
\eea
It is worth noting that for this case the form of behind the outer horizon cut off for Gauss-Bonnet-Maxwell theory (\ref{r0GB}) is the same as one for Einstein-Hilbert-Maxwell theory (\ref{r0EH}) but now $r_-$ is the location
of inner horizon for the black hole solution (\ref{key}) in Gauss-Bonnet-Maxwell theory. The general answer away from near extremality can also be obtained but it needs to simplify more the clutter of equations.
\section{Conclusions}
In this paper we have studied holographic complexity for charged AdS black holes at finite cut off in Einstein-Hilbert-Maxwell theory and Gauss-Bonnet-Maxwell theory. Our main motivation was to extend the analysis of \cite{Akhavan:2018wla},  which was done for neutral black branes at finite cut off, to the $U(1)$ charged black holes.\\

The key point in \cite{Akhavan:2018wla} is that the authors demanded that the growth rate of complexity at finite cut off and at late times is equal to two times of quasi local gravitational energy. This provision implies that a behind the horizon cut off exists in addition to the boundary cut off. For a $U(1)$ charged system, two different bounds are proposed for the late time behavior of complexity growth rate. In \cite{Brown:2015lvg} this bound is expressed as a special combination of physical charges, mass $M$ and the $U(1)$ charge $Q$. For small black holes with spherical horizon this bound is saturated and its value is $2(M-\mu Q)$. In \cite{Cai:2016xho} another bound on the growth rate of complexity at late times is proposed, which for any size of black hole it is given by subtraction of chemical potentials evaluated on outer and inner horizon, times the $U(1)$ charge. Interestingly, this second proposal implicitly implies that the growth rate of complexity in a CFT is equal to the $"T\bar{T}"$ deformation of that CFT. The reason is that the $"T\bar{T}"$ deformation of a CFT can be described by a geometry at boundary finite cut off but for any black hole size, no boundary cut off dependency appears in the second proposal. On the contrary, the first proposal at least for small black holes with spherical horizon implies that the complexity can be changed with the $"T\bar{T}"$ deformation. 
\\

In this work we study the consequences of assuming the first proposal. We see that in order
to have a late time behavior consistent with generalized Lloyd's bound \cite{Brown:2015lvg} one is forced to have a cut off behind the outer horizon and in front of inner horizon whose value is fixed by the inner horizon and boundary cut off. If one presumes that the second proposal is correct this cut off moves to behind the inner horizon in vicinity of curvature singularity \cite{Alishahiha:2019cib}. All these mean that the location of this new cut off crucially depends on whether the complexity of $U(1)$ charged CFTs changes by "$T\bar{T}$" deformations, or not. It might be a cancellation that happens between change of energy and change of chemical potantial times the total $U(1)$ charge, which causes that the complexity does not alter. This might be the case but it needs further exploration. One interesting way might be using the machinery which is developed in \cite{Jefferson:2017sdb}.\\  

It is worth mentioning that the result (\ref{r0GB}) is reliable for small value of $\alpha_{\mathrm{GB}}$ coupling. This is because in subsection.\ref{QLEGB} the proper counterterms (\ref{ct}) are acceptable just for small values of this coupling. It would be interesting to check the validity of (\ref{r0GB}) for arbitrary values of $\alpha_{\mathrm{GB}}$ coupling. Moreover, the results (\ref{r0EH}) and (\ref{r0GB}) are derived by the late time behavior of complexity growth rate and it would be interesting to find a time dependent behind the horizon cut off by using the full time dependency of holographic complexity.

\subsection*{Acknowledgements}

Special thank to M. Alishahiha for discussions and encouragements. The authors  would like also to kindly thank K. Babaei, A. Faraji Astaneh, M. R. Mohammadi
Mozaffar, F. Omidi, M. R. Tanhayi and M.H. Vahidinia for useful comments and discussions on related topics.


\begin{thebibliography}{99}


\bibitem{Ryu:2006bv} 
S.~Ryu and T.~Takayanagi,
``Holographic derivation of entanglement entropy from AdS/CFT,''
Phys.\ Rev.\ Lett.\  {\bf 96}, 181602 (2006)


\bibitem{Susskind:2014moa} 
L.~Susskind,
``Entanglement is not enough,''
Fortsch.\ Phys.\  {\bf 64}, 49 (2016)



\bibitem{Mathur:2009hf} 
S.~D.~Mathur,
``The Information paradox: A Pedagogical introduction,''
Class.\ Quant.\ Grav.\  {\bf 26}, 224001 (2009)



\bibitem{Almheiri:2012rt} 
A.~Almheiri, D.~Marolf, J.~Polchinski and J.~Sully,
``Black Holes: Complementarity or Firewalls?,''
JHEP {\bf 1302}, 062 (2013)



\bibitem{Almheiri:2013hfa} 
A.~Almheiri, D.~Marolf, J.~Polchinski, D.~Stanford and J.~Sully,
``An Apologia for Firewalls,''
JHEP {\bf 1309}, 018 (2013)


\bibitem{Marolf:2013dba} 
D.~Marolf and J.~Polchinski,
``Gauge/Gravity Duality and the Black Hole Interior,''
Phys.\ Rev.\ Lett.\  {\bf 111}, 171301 (2013)


\bibitem{Papadodimas:2012aq} 
K.~Papadodimas and S.~Raju,
``An Infalling Observer in AdS/CFT,''
JHEP {\bf 1310}, 212 (2013)


\bibitem{Papadodimas:2013jku} 
K.~Papadodimas and S.~Raju,
``State-Dependent Bulk-Boundary Maps and Black Hole Complementarity,''
Phys.\ Rev.\ D {\bf 89}, no. 8, 086010 (2014)


\bibitem{deBoer:2018ibj} 
J.~De Boer, S.~F.~Lokhande, E.~Verlinde, R.~Van Breukelen and K.~Papadodimas,
``On the interior geometry of a typical black hole microstate,''
arXiv:1804.10580 [hep-th].


\bibitem{Arora:2009} 
S.~Arora and B.~Barak,
``Computational complexity: A modern approach,''
Cambridge
University Press (2009).

\bibitem{Moore: 2011} 
C.~Moore,
``The nature of computation,''
Oxford University Press.

\bibitem{Alishahiha:2015rta} 
M.~Alishahiha,
``Holographic Complexity,''
Phys.\ Rev.\ D {\bf 92}, no. 12, 126009 (2015)


\bibitem{Brown:2015bva} 
A.~R.~Brown, D.~A.~Roberts, L.~Susskind, B.~Swingle and Y.~Zhao,
``Holographic Complexity Equals Bulk Action?,''
Phys.\ Rev.\ Lett.\  {\bf 116}, no. 19, 191301 (2016)


\bibitem{Brown:2015lvg} 
A.~R.~Brown, D.~A.~Roberts, L.~Susskind, B.~Swingle and Y.~Zhao,
``Complexity, action, and black holes,''
Phys.\ Rev.\ D {\bf 93}, no. 8, 086006 (2016)


\bibitem{Alishahiha:2018lfv} 
M.~Alishahiha, K.~Babaei Velni and M.~R.~Mohammadi Mozaffar,
``Subregion Action and Complexity,''
arXiv:1809.06031 [hep-th].



\bibitem{Lloyd} 
S.~Lloyd,
``Ultimate physical limits to computation,''
Nature 406 (2000) 1047, [arXiv:quantph/9908043].


\bibitem{Swingle:2017zcd} 
B.~Swingle and Y.~Wang,
``Holographic Complexity of Einstein-Maxwell-Dilaton Gravity,''
JHEP {\bf 1809}, 106 (2018)


\bibitem{Alishahiha:2018tep} 
M.~Alishahiha, A.~Faraji Astaneh, M.~R.~Mohammadi Mozaffar and A.~Mollabashi,
``Complexity Growth with Lifshitz Scaling and Hyperscaling Violation,''
JHEP {\bf 1807}, 042 (2018)


\bibitem{Lehner:2016vdi} 
L.~Lehner, R.~C.~Myers, E.~Poisson and R.~D.~Sorkin,
``Gravitational action with null boundaries,''
Phys.\ Rev.\ D {\bf 94}, no. 8, 084046 (2016)


\bibitem{Carmi:2017jqz} 
D.~Carmi, S.~Chapman, H.~Marrochio, R.~C.~Myers and S.~Sugishita,
``On the Time Dependence of Holographic Complexity,''
JHEP {\bf 1711}, 188 (2017)


\bibitem{Yang:2016awy} 
R.~Q.~Yang,
``Strong energy condition and complexity growth bound in holography,''
Phys.\ Rev.\ D {\bf 95}, no. 8, 086017 (2017)


\bibitem{Carmi:2016wjl} 
D.~Carmi, R.~C.~Myers and P.~Rath,
``Comments on Holographic Complexity,''
JHEP {\bf 1703}, 118 (2017)

\bibitem{Reynolds:2016rvl} 
A.~Reynolds and S.~F.~Ross,
``Divergences in Holographic Complexity,''
Class.\ Quant.\ Grav.\  {\bf 34}, no. 10, 105004 (2017)


\bibitem{Stanford:2014jda} 
D.~Stanford and L.~Susskind,
``Complexity and Shock Wave Geometries,''
Phys.\ Rev.\ D {\bf 90}, no. 12, 126007 (2014)


\bibitem{Chapman:2018dem} 
S.~Chapman, H.~Marrochio and R.~C.~Myers,
``Holographic complexity in Vaidya spacetimes. Part I,''
JHEP {\bf 1806}, 046 (2018)

\bibitem{Chapman:2018lsv} 
S.~Chapman, H.~Marrochio and R.~C.~Myers,
``Holographic complexity in Vaidya spacetimes. Part II,''
JHEP {\bf 1806}, 114 (2018)




\bibitem{Jefferson:2017sdb} 
R.~Jefferson and R.~C.~Myers,
``Circuit complexity in quantum field theory,''
JHEP {\bf 1710}, 107 (2017)


\bibitem{Chapman:2017rqy} 
S.~Chapman, M.~P.~Heller, H.~Marrochio and F.~Pastawski,
``Toward a Definition of Complexity for Quantum Field Theory States,''
Phys.\ Rev.\ Lett.\  {\bf 120}, no. 12, 121602 (2018)



\bibitem{Khan:2018rzm} 
R.~Khan, C.~Krishnan and S.~Sharma,
``Circuit Complexity in Fermionic Field Theory,''
Phys.\ Rev.\ D {\bf 98}, no. 12, 126001 (2018)



\bibitem{Hackl:2018ptj} 
L.~Hackl and R.~C.~Myers,
``Circuit complexity for free fermions,''
JHEP {\bf 1807}, 139 (2018)



\bibitem{Alves:2018qfv} 
D.~W.~F.~Alves and G.~Camilo,
``Evolution of complexity following a quantum quench in free field theory,''
JHEP {\bf 1806}, 029 (2018)


\bibitem{Camargo:2018eof} 
H.~A.~Camargo, P.~Caputa, D.~Das, M.~P.~Heller and R.~Jefferson,
``Complexity as a novel probe of quantum quenches: universal scalings and purifications,''
arXiv:1807.07075 [hep-th].



\bibitem{Caputa:2017urj} 
P.~Caputa, N.~Kundu, M.~Miyaji, T.~Takayanagi and K.~Watanabe,
``Anti-de Sitter Space from Optimization of Path Integrals in Conformal Field Theories,''
Phys.\ Rev.\ Lett.\  {\bf 119}, no. 7, 071602 (2017)


\bibitem{Sinamuli:2019utz} 
M.~Sinamuli and R.~B.~Mann,
``Holographic Complexity and Charged Scalar Fields,''
arXiv:1902.01912 [hep-th].


\bibitem{Bhattacharyya:2018bbv} 
A.~Bhattacharyya, A.~Shekar and A.~Sinha,
``Circuit complexity in interacting QFTs and RG flows,''
JHEP {\bf 1810}, 140 (2018)

\bibitem{Yang:2018tpo} 
R.~Q.~Yang, Y.~S.~An, C.~Niu, C.~Y.~Zhang and K.~Y.~Kim,
``More on complexity of operators in quantum field theory,''
arXiv:1809.06678 [hep-th].


\bibitem{Ali:2018fcz} 
T.~Ali, A.~Bhattacharyya, S.~Shajidul Haque, E.~H.~Kim and N.~Moynihan,
``Time Evolution of Complexity: A Critique of Three Methods,''
arXiv:1810.02734 [hep-th].


\bibitem{Zamolodchikov:2004ce} 
A.~B.~Zamolodchikov,
``Expectation value of composite field T anti-T in two-dimensional quantum field theory,''
hep-th/0401146.


\bibitem{Smirnov:2017} 
F.~Smirnov and A.~Zamolodchikov,
``On space of integrable quantum field theories,''
Nuclear Physics B 915, 363 – 383, 2017.


\bibitem{Cavaglia:2016oda} 
A.~Cavaglià, S.~Negro, I.~M.~Szécsényi and R.~Tateo,
``$T \bar{T}$-deformed 2D Quantum Field Theories,''
JHEP {\bf 1610}, 112 (2016).

\bibitem{McGough:2016lol} 
L.~McGough, M.~Mezei and H.~Verlinde,
``Moving the CFT into the bulk with $ T\overline{T} $,''
JHEP {\bf 1804}, 010 (2018)

\bibitem{Dubovsky:2017cnj} 
S.~Dubovsky, V.~Gorbenko and M.~Mirbabayi,
``Asymptotic fragility, near AdS$_{2}$ holography and $ T\overline{T} $,''
JHEP {\bf 1709}, 136 (2017)


\bibitem{Shyam:2017znq} 
V.~Shyam,
``Background independent holographic dual to $T\bar{T}$ deformed CFT with large central charge in 2 dimensions,''
JHEP {\bf 1710}, 108 (2017)


\bibitem{Kraus:2018xrn} 
P.~Kraus, J.~Liu and D.~Marolf,
``Cutoff AdS$_{3}$ versus the $ T\overline{T} $ deformation,''
JHEP {\bf 1807}, 027 (2018)



\bibitem{Cardy:2018sdv} 
J.~Cardy,
``The $ T\overline{T} $ deformation of quantum field theory as random geometry,''
JHEP {\bf 1810}, 186 (2018)


\bibitem{Aharony:2018vux} 
O.~Aharony and T.~Vaknin,
``The TT* deformation at large central charge,''
JHEP {\bf 1805}, 166 (2018)



\bibitem{Dubovsky:2018bmo} 
S.~Dubovsky, V.~Gorbenko and G.~Hernández-Chifflet,
``$ T\overline{T} $ partition function from topological gravity,''
JHEP {\bf 1809}, 158 (2018)


\bibitem{Taylor:2018xcy} 
M.~Taylor,
``TT deformations in general dimensions,''
arXiv:1805.10287 [hep-th].


\bibitem{Hartman:2018tkw} 
T.~Hartman, J.~Kruthoff, E.~Shaghoulian and A.~Tajdini,
``Holography at finite cutoff with a $T^2$ deformation,''
arXiv:1807.11401 [hep-th].


\bibitem{Akhavan:2018wla} 
A.~Akhavan, M.~Alishahiha, A.~Naseh and H.~Zolfi,
``Complexity and Behind the Horizon Cut Off,''
JHEP {\bf 1812}, 090 (2018)

%
%
%


\bibitem{Alishahiha:2018swh} 
M.~Alishahiha,
``On Complexity of Jackiw-Teitelboim Gravity,''
arXiv:1811.09028 [hep-th].

\bibitem{Alishahiha:2019cib} 
M.~Alishahiha, K.~Babaei Velni and M.~R.~Tanhayi,
``Complexity and Near Extremal Charged Black Branes,''
arXiv:1901.00689 [hep-th].


\bibitem{Jackiw:1984je} 
R.~Jackiw,
``Lower Dimensional Gravity,''
Nucl.\ Phys.\ B {\bf 252}, 343 (1985).

\bibitem{Teitelboim:1983ux} 
C.~Teitelboim,
``Gravitation and Hamiltonian Structure in Two Space-Time Dimensions,''
Phys.\ Lett.\  {\bf 126B}, 41 (1983).




\bibitem{Sachdev:92} 
Subir,~ Sachdev, Jinwu,~ Ye,
``Gapless Spin-Fluid Ground State in a Random Quantum Heisenberg Magnet,''
Phys. Rev. Lett. 70, 3339 (1993).


\bibitem{Kitaev:15} 
A.~Kitaev,
``A simple model of quantum holography,''
KITP strings seminar and Entanglement 2015 program
(Feb. 12, April 7, and May 27, 2015) . http://online.kitp.ucsb.edu/online/entangled15/.


\bibitem{Maldacena:2016upp} 
J.~Maldacena, D.~Stanford and Z.~Yang,
``Conformal symmetry and its breaking in two dimensional Nearly Anti-de-Sitter space,''
PTEP {\bf 2016}, no. 12, 12C104 (2016)


\bibitem{Brown:2018bms} 
A.~R.~Brown, H.~Gharibyan, H.~W.~Lin, L.~Susskind, L.~Thorlacius and Y.~Zhao,
``The Case of the Missing Gates: Complexity of Jackiw-Teitelboim Gravity,''
arXiv:1810.08741 [hep-th].


\bibitem{Goto:2018iay} 
K.~Goto, H.~Marrochio, R.~C.~Myers, L.~Queimada and B.~Yoshida,
``Holographic Complexity Equals Which Action?,''
arXiv:1901.00014 [hep-th].


\bibitem{Cai:2016xho} 
R.~G.~Cai, S.~M.~Ruan, S.~J.~Wang, R.~Q.~Yang and R.~H.~Peng,
``Action growth for AdS black holes,''
JHEP {\bf 1609}, 161 (2016)

\bibitem{Chamblin:1999tk} 
A.~Chamblin, R.~Emparan, C.~V.~Johnson and R.~C.~Myers,
``Charged AdS black holes and catastrophic holography,''
Phys.\ Rev.\ D {\bf 60}, 064018 (1999)

\bibitem{Parattu:2015gga} 
K.~Parattu, S.~Chakraborty, B.~R.~Majhi and T.~Padmanabhan,
``A Boundary Term for the Gravitational Action with Null Boundaries,''
Gen.\ Rel.\ Grav.\  {\bf 48}, no. 7, 94 (2016)


\bibitem{Parattu:2016trq} 
K.~Parattu, S.~Chakraborty and T.~Padmanabhan,
``Variational Principle for Gravity with Null and Non-null boundaries: A Unified Boundary Counter-term,''
Eur.\ Phys.\ J.\ C {\bf 76}, no. 3, 129 (2016)


\bibitem{Cano:2018aqi} 
P.~A.~Cano, R.~A.~Hennigar and H.~Marrochio,
``Complexity Growth Rate in Lovelock Gravity,''
Phys.\ Rev.\ Lett.\  {\bf 121}, no. 12, 121602 (2018)


\bibitem{Myers:1987yn} 
R.~C.~Myers,
``Higher Derivative Gravity, Surface Terms and String Theory,''
Phys.\ Rev.\ D {\bf 36}, 392 (1987).


\bibitem{Liu:2008zf} 
J.~T.~Liu and W.~A.~Sabra,
``Hamilton-Jacobi Counterterms for Einstein-Gauss-Bonnet Gravity,''
Class.\ Quant.\ Grav.\  {\bf 27}, 175014 (2010)

\bibitem{jafari2019} 
Ghadir.~Jafari,
``To be published.''.

\bibitem{Chakraborty:2018dvi} 
S.~Chakraborty and K.~Parattu,
``Null boundary terms for Lanczos-Lovelock gravity,''
arXiv:1806.08823 [gr-qc].


\end{thebibliography}
\end{document}